\documentclass[twocolumn,aps,prl,showpacs,superscriptaddress]{revtex4-1}
\usepackage{lmodern}

\usepackage[latin9]{inputenc} 
\usepackage{amsmath}
\usepackage{amssymb}
\usepackage{graphicx}
\usepackage{epstopdf}
\usepackage{enumerate}
\usepackage{ulem}
\usepackage{multirow}
\usepackage{xfrac}
\usepackage[usenames, dvipsnames]{color}

\usepackage{hyperref}
\hypersetup{colorlinks=true,linkcolor=black,citecolor=black}

\usepackage{xr}
\definecolor{mypink2}{RGB}{219, 48, 122}

\newcommand{\bb}[1]{\textbf{#1}}

\newcommand{\Teff}{$T_{\text{eff}}$}
\newcommand{\Tsat}{$T_{\text{sat}}$}
\newcommand{\Idc}{$I_{\text{DC}}$}
\newcommand{\Iac}{$I_{\text{AC}}$}
\newcommand{\nbse}{2H-NbSe\textsubscript{2}}
\newcommand{\s}{\space}
\newcommand{\ino}{a:InO}

\begin{document}
\bibliographystyle{naturemag}

\title{Extreme Sensitivity of the Superconducting State in Thin Films}

\author{I. Tamir}
\affiliation{Department of Condensed Matter Physics, The Weizmann Institute of Science, Rehovot 76100, Israel.}
\email{idan.tamir@weizmann.ac.il.; Corresponding author}
\author{A. Benyamini}
\affiliation{Department of Mechanical Engineering, Columbia University, New York, New York 10027, USA.}
\author{E. J. Telford} 
\affiliation{Department of Physics, Columbia University, New York, New York 10027, USA.}
\author{F. Gorniaczyk}
\affiliation{Department of Condensed Matter Physics, The Weizmann Institute of Science, Rehovot 76100, Israel.}
\author{A. Doron}
\affiliation{Department of Condensed Matter Physics, The Weizmann Institute of Science, Rehovot 76100, Israel.}
\author{T. Levinson}
\affiliation{Department of Condensed Matter Physics, The Weizmann Institute of Science, Rehovot 76100, Israel.}
\author{D. Wang} 
\affiliation{Department of Physics, Columbia University, New York, New York 10027, USA.}
\author{F. Gay}
\affiliation{Univ. Grenoble Alpes, CNRS, Grenoble INP, Institut N\'{e}el, 38000 Grenoble, France}
\author{B. Sac\'ep\'e}
\affiliation{Univ. Grenoble Alpes, CNRS, Grenoble INP, Institut N\'{e}el, 38000 Grenoble, France}	
\author{J. Hone}
\affiliation{Department of Mechanical Engineering, Columbia University, New York, New York 10027, USA.}
\author{K. Watanabe}
\affiliation{National Institute for Materials Science, 1-1 Namiki, Tsukuba, 305-0044 Japan.}
\author{T. Taniguchi}
\affiliation{National Institute for Materials Science, 1-1 Namiki, Tsukuba, 305-0044 Japan.}
\author{C. R. Dean}
\affiliation{Department of Physics, Columbia University, New York, New York 10027, USA.}
\author{A. N. Pasupathy}
\affiliation{Department of Physics, Columbia University, New York, New York 10027, USA.}
\author{D. Shahar}
\affiliation{Department of Condensed Matter Physics, The Weizmann Institute of Science, Rehovot 76100, Israel.}

\begin{abstract}

\end{abstract}

\maketitle

%
%
%
{\bf All non-interacting two-dimensional electronic systems are expected to exhibit an insulating ground state \cite{GangOf4}. This conspicuous absence of the metallic phase has been challenged only in the case of low-disorder, low density, semiconducting systems \cite{PhysRevB.50.8039} where strong interactions dominate the electronic state. 
Unexpectedly, over the last two decades, there have been multiple reports on the observation of a state with metallic characteristics on a variety of thin-film superconductors \cite{Ephronprl,PhysRevB.73.100505,crauste2009thickness,ye2012superconducting,PhysRevB.87.024509,saito2015metallic,tsen2015nature}.
To date, no theoretical explanation has been able to fully capture the existence of such a state for the large variety of superconductors exhibiting it. 
Here we show that for two very different thin-film superconductors, amorphous indium-oxide and a single-crystal of \nbse, this metallic state can be eliminated by filtering external radiation. Our results show that these superconducting films are extremely sensitive to external perturbations leading to the suppression of superconductivity and the appearance of temperature independent, metallic like, transport at low temperatures. We relate the extreme sensitivity to the theoretical observation that, in two-dimensions, superconductivity is only marginally stable \cite{fisherduality,FeigelmanPhysC}.}

It is theoretically accepted that in realistic two-dimensional systems, with the unavoidable disorder and at a finite $T$, superconductivity exists only marginally and finite resistivity is always expected \cite{fisherduality,FeigelmanPhysC}. The value of this residual resistance is extremely sensitive to the state of the system and it usually depends exponentially on experimental variables such as $T$, $B$, measurement current ($I$) and the level of microscopic disorder. An exception is the case of exactly zero magnetic field ($B$), which can effectively be attained in experiments, where true superconductivity, with $R\equiv 0$ at a finite $T$, is expected.

For such thin-film systems the superconducting state can be dramatically terminated with a transition to an insulating phase \cite{haviprl62,HebardPrl}.
In this superconductor-insulator transition (SIT), metallic behavior is expected to be restricted to an unstable point at the transition \cite{sondhirmp}.
This point of view is often supported by experiments using a variety of ways to drive the SIT including thickness variation, disorder, magnetic field ($B$) and carrier concentration (for a review, see \cite{physupekhi}).

There is, however, a growing number of independent studies \cite{Ephronprl,PhysRevB.73.100505,crauste2009thickness,ye2012superconducting,PhysRevB.87.024509,saito2015metallic,tsen2015nature} where the observation of an unexpected metallic state, intervening between the superconducting and insulating phases, has been reported. The unique characteristic attributed to this 'anomalous metal' is that the superconducting transition, signaled by an exponential decrease below a well-defined critical $T$ ($T_{C}$) of the sheet-resistance ($R$) from its normal state value ($R_N$) is terminated, upon further cooling, with a crossover to a $T$-independent $R$ that persists down to the lowest $T$'s.
This behavior, seen in thin films for which $R_N$ is significantly lower than the quantum of resistance $R_Q\equiv h/e^2\simeq25.8\space k\Omega$, is observed over a wide range of experimental parameters and extends to relatively high $T$'s \cite{kapitulnik2017anomalous}. Unlike ordinary metals this state exhibits a vanishing Hall-effect that was associated with a new particle-hole symmetric ground-state \cite{PhysRevB.93.205116,breznay2017particle}, its microwave response shows no cyclotron resonance and it reveals the existence of short-range superconducting correlations \cite{wang2017bose}.

The physical origin of this anomalous metallic state remains controversial with experimental measurements variously interpreted as evidence of a Bose-metal phase \cite{tsen2015nature} or dissipation arising from collective vortex tunneling \cite{Ephronprl,saito2015metallic}. 
Although several theoretical groups have addressed this state \cite{stauffer1994introduction,PhysRevLett.80.3352,PhysRevB.60.1261,PhysRevB.63.125322,phillips2003elusive,larkin2005theory,PhysRevLett.95.077002,PhysRevB.77.214523,PhysRevB.92.205104,PhysRevB.93.205116,PhysRevB.95.045118}, its robustness and ubiquitous nature pose difficulties in the development of a comprehensive model \cite{kapitulnik2017anomalous}. The purpose of this Letter is to show that the apparent metallic behavior can result from an unforeseen sensitivity of these marginal superconductors to external perturbations. 

\begin{figure*}[ht!]
	\includegraphics  [width=\textwidth] {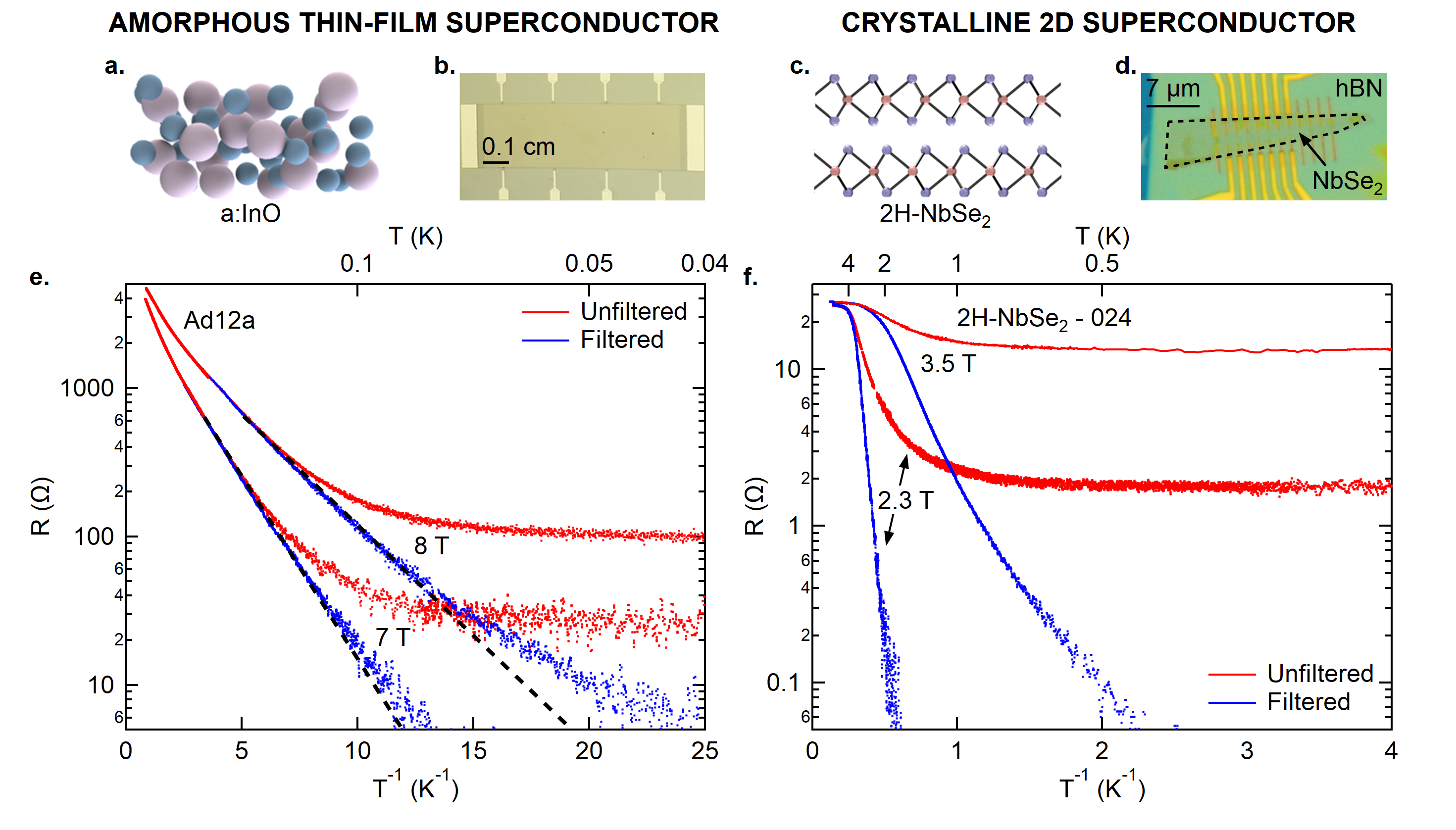}	
	\caption{\bb{Eliminating saturation.} \bb{a},\bb{c}, Illustration of the system structure and \bb{b},\bb{d}, microscope image of an a:InO (AD12a) and a \nbse\s (024) film respectively. \bb{e},\bb{f}, $R$ vs $T^{-1}$ obtained from an a:InO film at $B=7,8$ T, and a quad-layer \nbse\s at $B=2.3,3.5$ T respectively. Blue traces are measured with, and red traces without, filters. The top axis indicates the corresponding $T$'s. The black dashed lines in e, are guides to the eye indicating activated behavior. The data was measured applying a standard four-terminal lock-in technique with $I_0=1$ (a:InO) and 100 (\nbse) nA.
	}
	\label{filtercompare}
\end{figure*}

Our data were obtained from two very different superconducting systems. The first is amorphous Indium Oxide (a:InO) thin film (Fig. \ref{filtercompare}a,b), known for its high level of disorder reflected by high $R_N$ ($\lesssim \frac{h}{4e^2}\simeq 6.4$ k$\Omega$). The second system we investigated is made of sheets exfoliated from a single-crystal of \nbse\s (Fig. \ref{filtercompare}c,d), which are of high purity and are characterized by low $R_N$ ($< 100~\Omega$) \cite{telford2018via}.
Within the field of thin-film superconductors these two systems represent opposite limits with respect to structure and disorder. Moreover, while \nbse\s is a purely 2D superconductor having a thickness $d\ll\xi$, $\xi$ being the superconducting coherence length, $d$ of the a:InO films is approximately $5$ times larger than its $\xi$ \cite{PhysRevB.91.220508}. 

We begin by showing that the superconducting phase into which our samples transition at $T_C$, and which is interrupted as saturation sets in at lower $T$'s, is completely restored by introducing low-pass filters into the measurement setup (see Supplementary Fig. S1).
This is illustrated by plotting $R$ as a function of $T^{-1}$ obtained from an a:InO film (Fig. \ref{filtercompare}e) and a \nbse\s film (Fig. \ref{filtercompare}f). In both samples $R$ obtained from the unfiltered measurements (red traces) initially decreased exponentially with an approximate activated behavior $R(T)\propto exp(-U(B)/k_BT)$, where $U(B)$ is the activation energy and $k_B$ the Boltzmann constant. The exponential decrease then terminated with a transition to a saturated regime that persisted down to our lowest $T$'s. It is this saturated behavior of $R$ that was previously interpreted as indicating the novel metallic state \cite{Ephronprl,saito2015metallic,tsen2015nature,kapitulnik2017anomalous}.

\begin{figure*}[h]
	\includegraphics  [width=.5\textwidth] {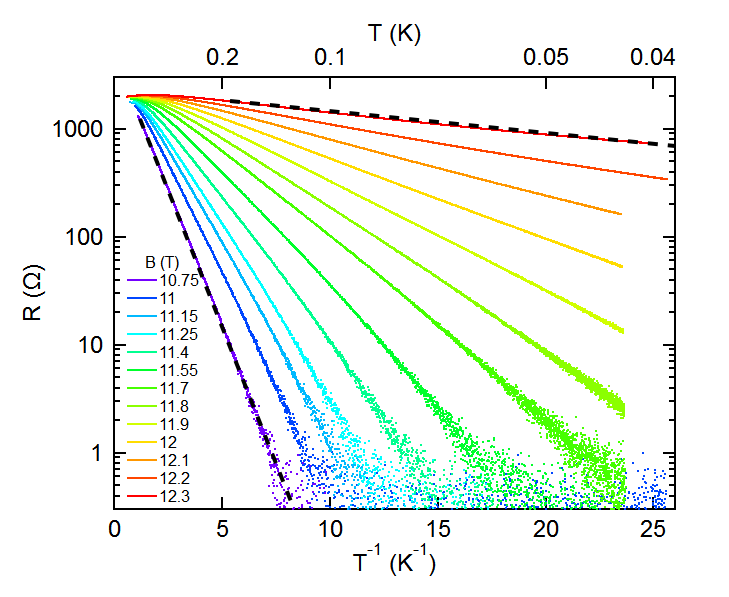}	
	\caption{\bb{Fully recovered activated behavior}. $R$ vs $T^{-1}$ obtained from IT1b20, an a:InO film, measured with better filtration (see Methods for details) at different $B$ values. Activated behavior, straight line in an Arrhenius plot, is apparent down to our noise floor or lowest measurement $T$'s, see \textit{e.g.} dashed black lines at $B=$10.75 and 12.3 T.}
	\label{activated}
\end{figure*}

When we repeated the measurements, this time with low-pass filters installed (blue traces), we found that $R$ continued to follow the activated trend down to much lower $T$'s, and, as $T$ was further lowered, $R$ continued to decrease to our noise level without saturating. The lowest $R$ we now measure can exceed two orders of magnitude below the corresponding saturated values of the unfiltered measurements. We note that with filtering, we continue to observe deviation from activated behavior in the lowest $T$ ranges measured. However, we believe this results from imperfect filters. Indeed, additional measurements of \ino\s film in a second fridge with improved low-$T$ filters show no deviation from activated behavior over the full range of achievable $T$'s (Fig. \ref{activated}). We conclude that our data do not support the existence of quantum corrections \cite{Ephronprl,saito2015metallic,PhysRevB.96.224511} to the well-known transport due to thermally activated vortices \cite{FeigelmanPhysC}.

\begin{figure*}[ht!]
	\includegraphics  [width=\textwidth] {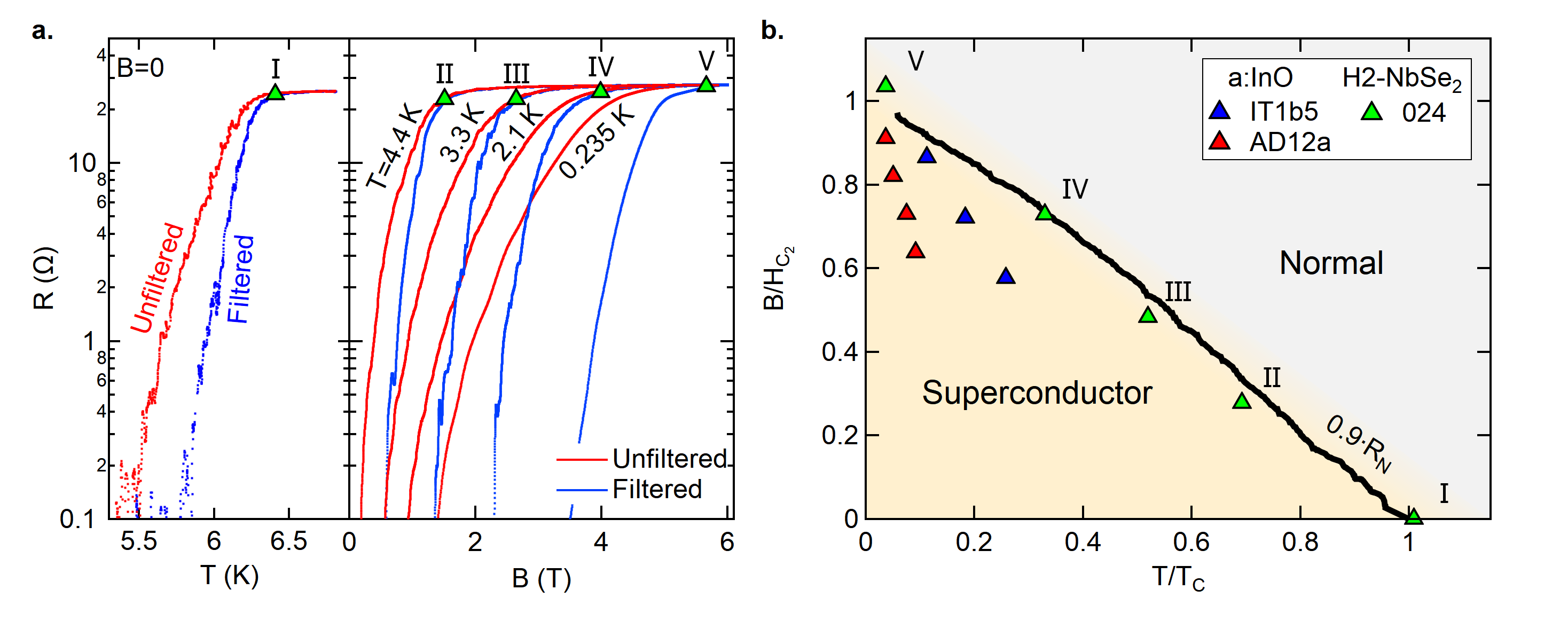}	
	\caption{\bb{Resistive Transition.} 
		\bb{a}, Measurements of the $T$ (left panel) and $B$ (right panel) driven superconducting-normal phase transition in a \nbse\s film (024). Blue traces are measured with, and red traces without, filters. \bb{b}, $B-T$ phase diagram. The black line separating the superconducting and normal phases, defined by $R$=$0.9\cdot R_N$, obtained from the \nbse\s sample, and includes data which were left out from \bb{a} for visibility. The $T$ and $B$ values where $\Delta R/R_F=(R_\text{Unfiltered}-R_\text{Filtered})/R_\text{Filtered}$=3\%, are marked, in both \bb{a} and \bb{b}, by green triangles. Both $T_C$ and $H_{C_2}$ are defined at $R$=0.9$\cdot R_N$. For our a:InO films, blue and red triangles, we used $T_C$=2.5 K (Ad12a) and 3 K (IT1b5).
	}
	\label{TB}
\end{figure*}

Although the effect the filters have on both systems is qualitatively similar it is important to point out that, while for \ino\s it is only seen well below $T_C$, for \nbse\s filtering has a measurable effect right from $T_C$, and at $B=0$. In Fig. \ref{TB}a we show the thermodynamic superconducting-normal transitions of \nbse, at $B=0$ in the left panel, and, near $H_{c2}$ for several $T$'s in the right. The common theme in these figures is that a significant effect of the filters is measured very close to the transition into superconductivity. In contrast, for \ino\s initial differences between filtered and unfiltered measurements are only seen much below $T_C$.
This is summarized in Fig. \ref{TB}b where we present the $B-T$ phase diagram for our samples. For \nbse\s the initial effect of the filters, indicated by green triangles, overlaps within error with the superconductor-normal phase boundary (defined by $R$=$0.9R_N$), while for \ino\s the filters significantly influence the results only well within the superconducting phase, blue and red triangles. 

\begin{figure*}[h]
	\includegraphics  [width=\textwidth] {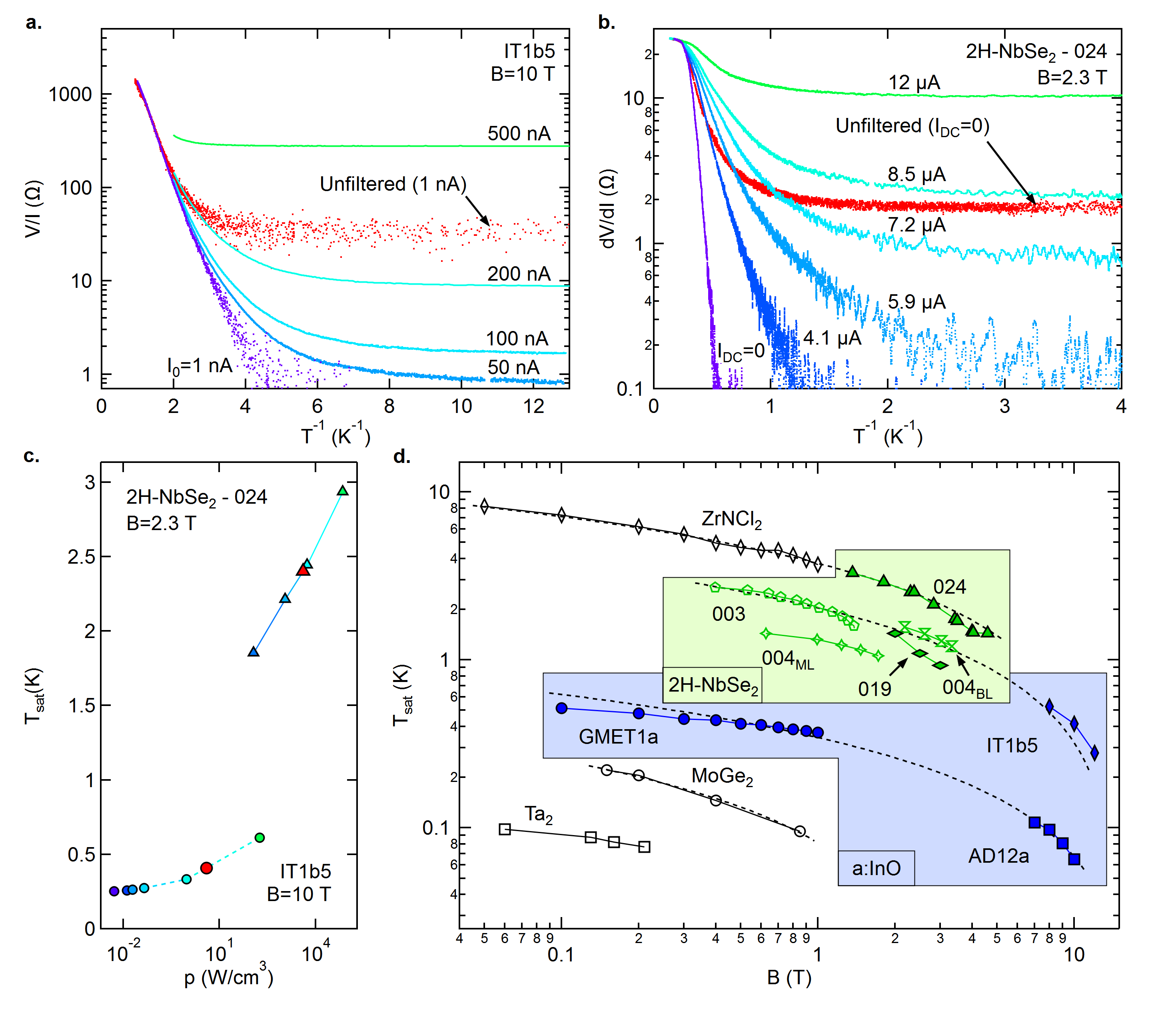}	
	\caption{\bb{Induced saturation and saturation T}. \bb{a}, $V/I$ vs $T^{-1}$ obtained from an a:InO film, measured at $B=10$ T with increasing $I_0$'s. 
		\bb{b}, $dV/dI$ vs $T^{-1}$ obtained from a \nbse\s film, measured at $B=2.3$ T with an increasing level of \Idc\s (see methods). In both \bb{a} and \bb{b}, we include one trace (red) measured, at our lowest $I$, without filters.
		\bb{c}, $T_{sat}^I$ vs. $p$ extracted from the a:InO data presented in \bb{a} (circles), and the \nbse\s data presented in \bb{b} (triangles), adopting the same color scale for the different $I$'s. The red symbols represent $T_{sat}^r$ extracted from the unfiltered measurements. The a:InO data include \Iac's left out of \bb{a} for visibility. These curves are used as $p$-meters to estimate $p_r=4$  and 4000 W/cm\textsuperscript{3} for the a:InO the \nbse\s sample, respectively.
		\bb{d.} \Tsat\s vs. $B$ evaluated for several samples. Our data is plotted in blue (a:InO) and green (\nbse). The MoGe \cite{Ephronprl}, Ta \cite{PhysRevB.73.100505}, ZrNCl \cite{saito2015metallic} data (black symbols) were extracted from the cited references. Full symbols represent data that were calculated using filtered measurements as a thermometer, while data plotted with empty symbols were estimated following the procedure in Ref. \cite{Ephronprl}. All samples exhibit similar logarithmic $B$ dependence, see dashed black lines.}
	\label{sat}
\end{figure*}

While filtering external radiation effectively eliminates the apparent metallic behavior, we found that saturation can be re-introduced by increasing the current used in our four-terminal measurements. For this purpose we used both DC and AC currents (\Idc\s and \Iac, see Methods) with similar results. The saturation induced by increasing \Iac\s is demonstrated in Fig.~\ref{sat}a, where we present data obtained from an a:InO film measured with filters at $B=10$ T, using increasing levels of $I_0$ (the amplitude of \Iac). While at $I_0=1$ nA the measured $V/I$, where $V$ is the voltage drop along the sample, followed an activated behavior with deviations that are barely noticeable over our noise level, for $I_0\geq 50$ nA the data significantly deviated from its low-$I_0$ value and saturation set-in at low $T$'s with the saturated value increasing with $I_0$. For reference we include one trace (red) measured without filters and at $I_0=1$ nA, which exhibits the low-$T$ saturation. We note that, since $R$ is an equilibrium value defined by $\lim_{I\rightarrow 0}V/I$, the data presented in Fig. \ref{sat}a strictly equals $R$ only in the Ohmic regime ($I_0\lesssim 1$ nA). Similar results are obtained while increasing \Idc\s, see Fig. \ref{sat}b where we present data obtained from a \nbse\s film.

The saturation induced by increasing $I$ (either AC or DC) can naturally be attributed to Joule heating. Under the application of a higher power ($P=I\cdot V$) by the measurement circuit, the electronic system is unable to equilibrate with its low-$T$ environment. This leads to an out-of-equilibrium steady-state where the electrons are held at an elevated $T$, \Teff, higher than the surrounding $T$ \cite{borisprl}. The $I$-induced deviations from activated behavior, as well as the saturation regions, can therefore be attributed to \Teff$>T$. We can, self-consistently, extract these \Teff's by fitting the $R(T)$ data, obtained from the filtered measurements in the Ohmic regime, with an activated form and then using this fit as our thermometry calibration curve: for each value of $V/I$, in the elevated-$I$ measurements, we associate a \Teff\s corresponding to $V/I=R$ in the calibration curve. Using this procedure we conveniently define \Tsat\s as \Teff\s in the $R$ saturation regime. 

We now wish to suggest that the $R$ saturation observed in our unfiltered experiments, and which bears a striking resemblance to the $I$-driven saturation (see Fig \ref{sat}a,b), can also be associated with heating. While in this case the source of heating is less obvious, the fact that filtering the electrical lines connected to the sample effectively eliminates the saturation suggests that the culprit is ambient radiation that propagates down the lines and couples directly to the low-$T$ electronic system. 
We can estimate the power density delivered to the electronic system by this radiation ($p_r$) by comparing \Tsat\s obtained during the application of known power density ($p$) in our filtered, elevated-$I$, measurements ($T_{sat}^I$) to the \Tsat\s obtained in the unfiltered, Ohmic, measurements ($T_{sat}^r$). To do this we plot, in Fig.~\ref{sat}c, the $p$ dependence of $T_{sat}^I$ for two of our samples and use it as our $p$-meter. Red diamonds indicate $T_{sat}^r$ values corresponding to the unfiltered curves of each sample.
We find $p_r=4$  and 4000 W/cm\textsuperscript{3} for the a:InO the \nbse\s samples, respectively. 


Although our results clearly show that our a:InO and \nbse\s films do not exhibit an intermediate metallic phase, we can not rule out the existence of such a phase in other superconducting systems for which a metallic state was previously reported \cite{Ephronprl,saito2015metallic}. We can, however, naively extended our effective-temperature analysis to these systems. In Fig.~\ref{sat}d we present \Tsat\s vs. $B$ obtained from our data (blue and green symbols), together with \Tsat\s values that we extracted from published data (black symbols). Whenever comparisons between results obtained from filtered and unfiltered measurements are not available (empty symbols), we fitted the data measured at higher $T$'s with activated behavior and used these fits as our thermometry calibration curves. This procedure, introduced in this context in Ref. \cite{Ephronprl}, only provides a lower bound for \Tsat\s because the filtered measurements can also exhibit higher $U(B)$ (Supplementary Fig. S2). While the data in Fig.~\ref{sat}d represent several very different systems, measured over a wide range of $T$'s, they all share a similar $B$ dependence: we found that \Tsat $\sim log(H_{C_2}/\alpha B)$, where $H_{C_2}$ is the upper critical field terminating superconductivity and $\alpha$ is a fit parameter of order 1, works reasonably well. 

Before we proceed to discuss the implication of this simple elevated effective-temperature scenario, we wish to point out that there are other possible mechanisms that would lead to $I$-dependent transport in thin-film superconductor such as ours. Non-linear vortex related response may be relevant at finite-$B$'s, and BKT vortex-antivortex unbinding may be at work at $B=0$. At this stage we are unable to rule out that such mechanisms play a significant role in the $I$-response of the system, and may even lead to saturated, $T$-independent $R$ as $T\rightarrow 0$. We are not aware of a model that accounts for the stark difference between the results of the filtered and unfiltered measurements.

The data we are presenting here show that the metallic behavior, often observed in thin-film superconductors, results from the exposure of the superconducting phase to unwanted radiation or high $I$'s. While these can be easily eliminated, it is still worthwhile to consider why these superconductors so readily respond to excitations that leave other systems, under similar conditions, unaffected. 
We point out that \Tsat\s is routinely around a few K, where it is unlikely that the cryogenic environment will limit the sample's ability to cool. Since the external power couples only to the electronic system it is reasonable to conclude that the bottleneck in the heat-transfer process is between the electrons and the host phonons \cite{borisprl}. This is not surprising since in superconductors the electron condensate is decoupled from the heat-carrying phonons. If such a limiting mechanism is at play a much more thorough theoretical analysis is necessary before we can go any further with quantitative tests.

In this study we were able to compare two very different systems under virtually identical measurement conditions. It is reasonable to assume that, without filters, the radiation delivered to both types of samples would be the same. Surprisingly we find that their \Tsat\s are very different: $<0.4$ K for \ino\s and $\sim2$ K for \nbse. Similarly the corresponding $p_r$'s are different. If the effective-temperature picture is correct, we need to understand why two samples under similar external radiation end up responding in such a different manner. The reason, we believe, is rooted in the energy balance maintained by the electrons. Even if the radiation is the same, it is very likely that different systems will absorb this energy in different ways reflecting the specific details of their electronic state. Compounding this are possible differences between the strength of the coupling to the phonon system to which the electrons can transfer the energy absorbed from the radiation. A detailed understanding of this scenario awaits further theoretical developments.

In summary, 
we showed that two very different thin-film superconductors are extremely sensitive to external perturbations, and exhibit metallic-like, saturated, $T$-dependence in response to such perturbations. We suggested two possible mechanisms, one based on vortex depinning and in the other we assume that an overheated state exists where the electronic system is unable to equilibrate with its surrounding. In the latter case one should theoretically address not only the external power dissipated but also the heat flow away from the electronic system.

\section*{Methods}
In this work we studied several different a:InO and \nbse\s samples. Their relevant parameters are presented in Table S1 of the Supplementary information. Details of the growth and fabrication were previously published, see Ref. \cite{telford2018via} for \nbse\s, and Ref. \cite{Shaharprb} for a:InO.

Our data was measured utilizing a standard four-terminal lock-in technique, see sketch in Supplementary Fig. S1. A low-frequency ($f\sim 10$ Hz) AC voltage ($V_{\text{AC}}$) generated by a lock-in oscillator is transformed into $I$ using a large resistor ($R_s=$ 10-1000 M$\Omega$). This results in a probing current of the following form: $I_{\text{AC}}=I_0\cdot\cos(2\pi ft)$, where $t$ is time and $I_0=|V_{\text{AC}}|/R_s$. At low enough $I_0$'s, where the systems is in its Ohmic regime, $R=V/I_{\text{AC}}$, where $V$ is measured at the same AC frequency along the sample. On to this AC signal we can superimpose a DC bias $I$ (\Idc). The total current in such setup is: $I_{tot}=I_{\text{AC}}+I_{\text{DC}}$. Since the lock-in measures only the AC component $V/$\Iac\s, in this configuration, is a measurement of the differential resistance.

The data presented in Fig. \ref{activated} were measured in a dilution refrigerator equipped with heavily filtered DC lines comprising feedthrough pi-filters at room $T$, low resistance twisted-pairs ($\sim$8 $\Omega$) from 300 K to 4 K to reduce Johnson noise, lossy shielded twisted-pairs ($\sim$500 $\Omega$) from 4 K to the mixing chamber (MC) stage, copper-powder filter \cite{PhysRevB.35.4682} on the MC stage, and cryogenic-compatible 47 nF capacitor-to-ground on the sample holder. Furthermore, as thermometry below 0.1 K is delicate, a RuO\textsubscript{2} thermometer calibrated by 60Co nuclear orientation thermometry has been installed on the sample holder close to the sample, thus subjected to the same cooling power from the wiring. This additional on-chip thermometry suppresses the very last deviations from activated transport at the lowest $T$'s.

\section*{Acknowledgments}
We are grateful to I. Aleiner, B. Altshuler, M. Feigelman, D. Kennes, S.A. Kivelson, K. Michaeli, A. Millis, P.W. Phillips, and B. Spivak for fruitful discussions.
This research was supported by The Israel Science Foundation (ISF Grant no. 556/17), the Minerva Foundation with funding from the Federal German Ministry for Education ad Research, the NSF MRSEC program through Columbia in the Center for Precision Assembly of Superstratic and Superatomic Solids (DMR-1420634), the Global Research Laboratory (GRL) Program (2016K1A1A2912707) funded by the Ministry of Science, ICT and Future Planning via the National Research Foundation of Korea (NRF), and Honda Research Institute USA Inc. 

\section*{Author contributions}
All authors participated in all aspects of this work.
\section*{Competing financial interests}
The authors declare no competing financial interests.

\bibliography{../s1ahir}

\end{document}


\bibliographystyle{apsrev}

\title{Supplementary Information of the Letter - Extreme Sensitivity of the Superconducting State in Thin Films}

\author{I. Tamir}
\affiliation{Department of Condensed Matter Physics, The Weizmann Institute of Science, Rehovot 76100, Israel.}
\email{idan.tamir@weizmann.ac.il.; Corresponding author}
\author{A. Benyamini}
\affiliation{Department of Mechanical Engineering, Columbia University, New York, New York 10027, USA.}
\author{E. J. Telford} 
\affiliation{Department of Physics, Columbia University, New York, NY, USA.}
\author{F. Gorniaczyk}
\affiliation{Department of Condensed Matter Physics, The Weizmann Institute of Science, Rehovot 76100, Israel.}
\author{A. Doron}
\affiliation{Department of Condensed Matter Physics, The Weizmann Institute of Science, Rehovot 76100, Israel.}
\author{T. Levinson}
\affiliation{Department of Condensed Matter Physics, The Weizmann Institute of Science, Rehovot 76100, Israel.}
\author{Da Wang} 
\affiliation{Department of Physics, Columbia University, New York, NY, USA.}
\author{F. Gay}
\affiliation{Univ. Grenoble Alpes, CNRS, Grenoble INP, Institut N\'{e}el, 38000 Grenoble, France}
\author{B. Sac\'ep\'e}
\affiliation{Univ. Grenoble Alpes, CNRS, Grenoble INP, Institut N\'{e}el, 38000 Grenoble, France}	
\author{J. Hone}
\author{K. Watanabe}
\affiliation{National Institute for Materials Science, 1-1 Namiki, Tsukuba, 305-0044 Japan.}
\author{T. Taniguchi}
\affiliation{National Institute for Materials Science, 1-1 Namiki, Tsukuba, 305-0044 Japan.}
\affiliation{Department of Mechanical Engineering, Columbia University, New York, New York 10027, USA.}
\author{C. R. Dean}
\affiliation{Department of Physics, Columbia University, New York, NY, USA.}
\author{A. N. Pasupathy}
\affiliation{Department of Physics, Columbia University, New York, NY, USA.}
\author{D. Shahar}
\affiliation{Department of Condensed Matter Physics, The Weizmann Institute of Science, Rehovot 76100, Israel.}

\maketitle

\renewcommand{\thefigure}{S\arabic{figure}}
\renewcommand{\thetable}{S\arabic{table}}

\begin{figure}[ht!]
	\includegraphics  [width=0.5\textwidth] {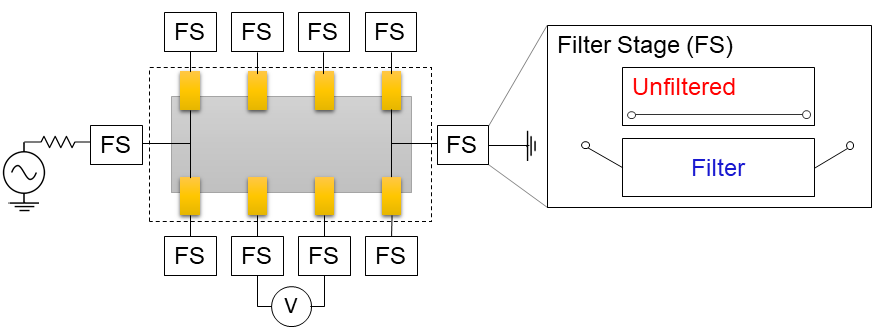}	
	\caption{\bb{Schematics of the measurement circuit}. The filter illustrated in the filter stage (FS) represent either an homemade RC filter, a commercial low-pass (Pi) filter, or an attenuator, all with similar effect. The dashed black square represents our cryogenic system. 
	}
	\label{config}
\end{figure}

\begin{figure}[ht!]
	\includegraphics  [width=0.5\textwidth] {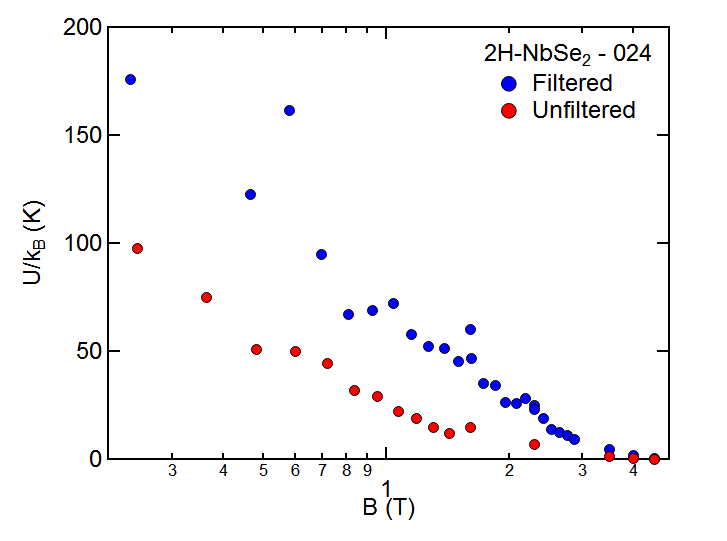}	
	\caption{\bb{Activation energy}. $U$, normalized by $k_B$, vs. $B$ obtained from a \nbse\s sample measured with (blue) and without (red) filters.}
	\label{UofB}
\end{figure}

\begin{table*}[ht!]
\begin{tabular} {l| l c c c c c c }
\hline
\hline
\rule{0pt}{12pt}System&Name&\multirow{2}{*}{\shortstack{$R_N$\\($\hbar/e^2$)}}& \multirow{2}{*}{\shortstack{$T_C$\\(K)}} & \multirow{2}{*}{\shortstack{$H_{C_2}$\\(T)}} & \multirow{2}{*}{\shortstack{Width\\($\mu$m)}}& \multirow{2}{*}{\shortstack{Thickness (nm)\\(\# of layers)}} &Structure\\
\\
\hline
a:InO & IT1b5 & 0.07 & $>2$ & 13.9 & 5 & 30 & \multirow{4}{*}{amorphous} \\
 & IT1b20 & 0.09 & $>2$ & 12.95 & 20 & 30 &  \\
 & GMET1 & 0.16 & $>2$ & $>12$ & 100 & 30 &  \\
 & AD12a & 0.23 & $>2$ & 11 & 333 & 22 &  \\
\hline
\rule{0pt}{12pt}
\nbse & 003 & $1.4\cdot10^{-3}$ & 5.7 & 4.67 & 1 & 2 & \multirow{5}{*}{\shortstack{single\\crystal}} \\
 & 004\textsubscript{ML} & $6.9\cdot10^{-4}$ & 4.3 & 3.2 & 1 & 1 &  \\
 & 004\textsubscript{BL} & $7.4\cdot10^{-4}$ & 5.86 & 4.38 & 1 & 2 & \\
 & 019 & $1.3\cdot10^{-3}$ & 6 & 4.12 & 3 & 2 &\\
 & 024 & $9.7\cdot10^{-4}$ & 6.35 & 5.48 & 1.2 & 4 & \\
\hline
\hline
\end{tabular}
\caption{\textbf{Sample parameters}. estimated parameters of the different samples measured in this work. For the \nbse\s films we note the number of layers, rather than thickness, in the 7\textsuperscript{th} column. The uncertainties in the a:InO films parameters is due to our setup limited $T$ and $B$ ranges.}
\label{SampleParameters}
\end{table*}

\bibliography{../../s1ahir}